\documentclass[prl,twocolumn,showpacs,floatfix,amsmath,amssymb,superscriptaddress]{revtex4}
\usepackage[dvips]{graphicx}
\usepackage{latexsym}
\usepackage{graphicx}
\usepackage{times}
\usepackage{amsmath}
\usepackage{dcolumn}
\usepackage{latexsym,amsmath,amssymb,bm,euscript}
\begin{document}

\title{Interlayer coherent composite Fermi liquid phase in quantum Hall bilayers}

\author{Jason Alicea}
\affiliation{Department of Physics, California Institute of Technology,
Pasadena, California 91125}

\author{Olexei I. Motrunich}
\affiliation{Department of Physics, California Institute of Technology,
Pasadena, California 91125}

\author{G. Refael}
\affiliation{Department of Physics, California Institute of Technology,
Pasadena, California 91125}

\author{Matthew P. A. Fisher}
\affiliation{
Microsoft Research, Station Q, University of California,
Santa Barbara, California 93106}

\date{\today}

\begin{abstract}
Composite fermions have played a seminal role in understanding the quantum Hall effect,
particularly the formation of a compressible `composite Fermi liquid' (CFL) at filling factor 
$\nu = 1/2$.  Here we suggest that in multi-layer systems interlayer Coulomb 
repulsion can similarly generate `metallic' behavior of composite fermions \emph{between} layers, 
even if the electrons remain insulating.   Specifically, we propose that a quantum Hall 
bilayer with $\nu = 1/2$ per layer at intermediate layer separation may host such an 
\emph{interlayer coherent CFL}, driven by exciton condensation of composite fermions.  This phase has 
a number of remarkable properties: the presence of `bonding' and `antibonding' composite Fermi 
seas, compressible behavior with respect to symmetric currents, and fractional quantum Hall behavior 
in the counterflow channel.  Quantum oscillations associated with the Fermi seas give rise to a new
series of incompressible states at fillings $\nu = p/[2(p\pm 1)]$ per layer ($p$ an integer), 
which is a bilayer analogue of the Jain sequence.  
\end{abstract}

\maketitle


Composite fermions have played a central role in the field of quantum 
Hall physics \cite{JainCF}.  Perhaps the most striking manifestation of composite fermions
is their formation of a \emph{Fermi sea} 
at certain even-denominator fillings, notably $\nu = 1/2$.  Pioneering work by Halperin, Lee, and 
Read (HLR) developed the theory of such composite Fermi liquids (CFL's) \cite{HLR}, and the 
anticipated Fermi surface has been experimentally measured \cite{CFbook}.  As a corollary of this interaction-driven `metallicity', CFL's also provide a unified picture for the onset of Jain's sequence \cite{JainCF}---these quantum 
Hall states emerge as quantum oscillations of a composite Fermi sea \cite{HLR}.  

Given the success of HLR theory, it is natural to inquire whether CFL's 
can emerge in strongly coupled {\it multi-layer} systems.  More precisely, can interlayer Coulomb repulsion generate coherent propagation of composite fermions \emph{between} 
layers, resulting in an \emph{interlayer coherent CFL} with a higher-dimensional composite 
Fermi surface?  Such a phase would constitute a fundamentally new kind of CFL and, if found, 
would broaden the utility of composite fermions into a new dimension.  Experimentally, this question 
is motivated in part by quantum Hall bilayers, for which compressible states appear even when 
interlayer Coulomb is `strong' (\emph{e.g.}, \cite{JimDrag}).  Additionally, recent experiments on bismuth \cite{Behnia} 
highlight our lack 
of understanding of strongly interacting three-dimensional systems in the lowest Landau level 
(LLL), further stimulating the quest for exotic multi-layer phases.  

\begin{figure}
\centering{
  \includegraphics[width=2.5in]{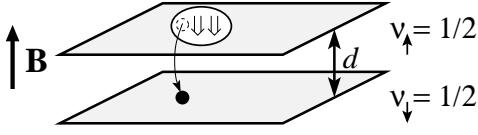}
  \caption{Schematic of bilayer setup.  
  The physical electron is a bound state of 2 flux quanta and 
  a composite fermion (shown by the small circles).  In the interlayer coherent CFL, composite 
  fermions tunnel between layers, while the electrons do not.  }
  \label{BilayerFig}}
\end{figure}

\begin{figure}
\centering{
  \includegraphics[width=3in]{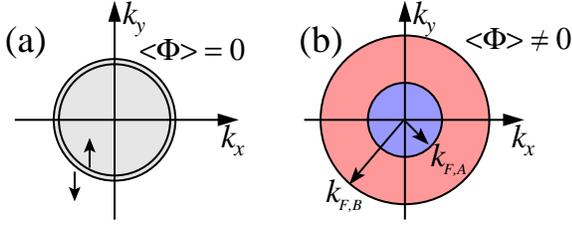}
  \caption{(Color online) (a) At $d/\ell_B = \infty$ and $\nu = 1/2$ per layer, composite fermions 
  form decoupled CFL's with identical Fermi surfaces in the $\uparrow$ and $\downarrow$ layers.  
  (b) At intermediate $d/\ell_B$, we propose an interlayer coherent CFL where composite 
  fermions coherently tunnel between layers and form bonding and antibonding Fermi surfaces 
  with radii $k_{F,B}$ and $k_{F,A}$.  }
  \label{FermiSurfaces}}
\end{figure}

Here we argue that spin-polarized quantum Hall bilayers at $\nu = 1/2$ per layer may 
indeed harbor an interlayer coherent CFL when the layer spacing $d$ and magnetic length
$\ell_B$ are comparable 
(see Fig.\ \ref{BilayerFig}).  To motivate this phase, it is useful to recall the well-understood physics at extreme $d/\ell_B$.  
For $d/\ell_B \lesssim 1$, strong interlayer Coulomb drives exciton condensation of the electrons \cite{QHbook}, $\langle c^\dagger_\uparrow c_\downarrow \rangle \neq 0$, with $c_{\alpha}$ the electron operator in layer
$\alpha = \uparrow, \downarrow$.
When $d/\ell_B \rightarrow \infty$, the 
layers decouple and form $\nu = 1/2$ CFL's with independent Fermi surfaces in each layer as in 
Fig.\ \ref{FermiSurfaces}(a).  Here the bilayer wavefunction is 
$\psi_{\infty} = P_{LLL}\prod_{i<j}(z_i-z_j)^2(w_i-w_j)^2 \Psi_{CF}^\uparrow \Psi_{CF}^\downarrow$,
where $P_{LLL}$ denotes LLL projection, $z,w$ are complex coordinates in layer 
$\uparrow,\downarrow$, and $\Psi_{CF}^\alpha$ is the composite Fermi sea wavefunction in layer $\alpha$.  The behavior at intermediate $d/\ell_B$ is 
far subtler and has been actively studied for more than a decade (for a recent discussion, 
see \cite{Moller}).  

We propose that for intermediate $d/\ell_B$, the short-range part of interlayer Coulomb 
naturally favors exciton condensation of \emph{composite fermions}, and that this leads to 
an interlayer coherent CFL.  Denoting the composite fermion operator by $f_\alpha$, this phase 
is characterized by $\langle f^\dagger_\uparrow f_\downarrow \rangle \neq 0$ even though 
$\langle c^\dagger_\uparrow c_\downarrow \rangle = 0$.   This order parameter implies
that composite fermions spontaneously 
tunnel between layers (even though the electrons do not), 
resulting in the formation of bonding and antibonding composite Fermi surfaces as shown in 
Fig.\ \ref{FermiSurfaces}(b).  A simple trial wavefunction for this new phase is
\begin{equation}
  \psi = P_{LLL}\prod_{i<j}(z_i-z_j)^2(w_i-w_j)^2 \Psi_{(k_{F,B},k_{F,A})} ,
  \label{coherentCFLpsi}
\end{equation}
where $k_{F,B/A}$ are the Fermi momenta for the bonding/antibonding Fermi surfaces and
$\Psi_{(k_{F,B},k_{F,A})}$ denotes the Slater determinant filling these Fermi seas.  While the 
interlayer coherent CFL behaves similarly to decoupled CFL's in response to symmetric currents, 
this phase has the remarkable property that \emph{in the counterflow channel it behaves as 
an incompressible $\nu = 1/2$ quantum Hall state}.  This follows from composite fermion 
exciton condensation, just as electron exciton condensation leads to counterflow superfluidity at 
small $d/\ell_B$ \cite{QHbook}.  Interestingly, quantum oscillations of the Fermi surfaces 
generate a bilayer analogue of Jain's sequence [see Eqs.\ (\ref{nu}) and (\ref{psipApB})], 
which includes Halperin's 331 state \cite{mmn} and other fractions that have been experimentally observed.  

To flesh out this picture we consider spin-polarized electrons, in the idealized limit of 
zero interlayer tunneling.  
The appropriate 
Euclidean action is 
\begin{equation}
  S = \int_{x}\sum_{\alpha = \uparrow,\downarrow}c_{x\alpha}^\dagger \bigg{[}\partial_\tau  -\frac{(\nabla+ie {\bf A})^2}{2m}\bigg{]} c_{x\alpha} + S_{Coul},
  \label{Selectron}
\end{equation}
where $x = ({\bf r},\tau)$, 
${\bf B} = \nabla \times {\bf A}$ is the external field, and 
$S_{Coul} = S_{Coul}^{\uparrow\uparrow} + S_{Coul}^{\downarrow\downarrow} + S_{Coul}^{\uparrow\downarrow}$ encodes the Coulomb repulsion.  (Throughout we set $\hbar = c = 1$.)  We focus primarily on fillings $\nu_\uparrow = \nu_\downarrow = 1/2$ at intermediate 
$d/\ell_B$, sufficiently large that the exciton condensate is destroyed but small enough that 
interlayer Coulomb is not weak.  Although far from the $d/\ell_B = \infty$ limit, 
we postulate that composite fermions remain the 
`correct' degrees of freedom in this regime.  
Denoting
the Chern-Simons fields by $a_\alpha^\mu$, the action then becomes
\begin{eqnarray}
  S_{CF} &=& \int_{x}\sum_{\alpha = \uparrow,\downarrow}\bigg{\{}f_{x\alpha}^\dagger \bigg{[}(\partial_\tau-ia_\alpha^0)  - \frac{(\nabla -i {\bf a}_\alpha)^2}{2m^*}\bigg{]} f_{x\alpha} 
  \nonumber \\
  &+& \frac{i}{8\pi}(a^\mu_\alpha+e\mathcal{A}_\alpha^\mu) \epsilon_{\mu\nu\lambda}\partial_\nu (a_\alpha^\lambda+ e\mathcal{A}^\lambda_\alpha)\bigg{\}}  
  + S_{Coul}.
  \label{SCF}
\end{eqnarray}
Here $\mathcal{A}_\alpha^\mu = A^\mu + \delta A^\mu_\alpha$, with $\delta A^\mu_\alpha$ a probe field added for computing response properties below.  The Chern-Simons term on the 
second line attaches two flux quanta to each composite fermion, recovering the 
physical electron as shown schematically in Fig.\ \ref{BilayerFig}.  We allow the
composite fermion mass $m^*$ to differ from the bare electron mass $m$, since the two 
are unrelated when Landau level mixing is ignored \cite{HLR}.  

Equation (\ref{SCF}) was previously studied in important work by Bonesteel \emph{et al}.\
\cite{BonesteelNayak}  By examining the effect of \emph{long-wavelength} gauge fluctuations at 
$d/\ell_B \gg 1$, these authors argued that such coupled CFL's should generically undergo an 
interlayer BCS pairing instability.  This is rather natural at large $d/\ell_B$ 
within the dipole picture of decoupled CFL's \cite{DipoleBCSpicture}.  However, when $d/\ell_B$ is 
of order unity---which is our focus here---the layers are strongly coupled, so in this case one 
should first attack the problem by satisfying the \emph{short-distance, high-energy} physics.  This 
is our objective.

To this end, we focus on the interlayer Coulomb $S_{Coul}^{\uparrow\downarrow}$, decomposed 
into short-range and long-range pieces via $S_{Coul}^{\uparrow\downarrow} = S^{\uparrow\downarrow}_{sr} + S^{\uparrow\downarrow}_{lr}$.  The short-range part can be written
\begin{eqnarray}
  S^{\uparrow\downarrow}_{sr} &=& u\int_{x}f^\dagger_{x\uparrow}f^\dagger_{x\downarrow}f_{x\downarrow}f_{x\uparrow} = -u\int_{x}f^\dagger_{x\uparrow}f_{x\downarrow}f^\dagger_{x\downarrow}f_{x\uparrow};
  \label{Ssr}
\end{eqnarray}
including here interactions out to a range $\ell_B$, we crudely estimate 
$u \approx (e^2/d)(\pi \ell_B^2)$.  \emph{Short-range} interlayer Coulomb is thus clearly attractive 
in the particle-hole rather than the Cooper channel, and favors exciton condensation of 
composite fermions rather than BCS pairing.  To expose this competing excitonic instability, we 
decouple Eq.\ (\ref{Ssr}) with a Hubbard-Stratonovich field $\Phi$, which can be regarded as a 
composite fermion exciton condensate order parameter:
\begin{eqnarray}
  S^{\uparrow\downarrow}_{sr} &\rightarrow& \int_{x}\bigg{[}\frac{1}{u}|\Phi(x)|^2-(f_{x\uparrow}^\dagger f_{x\downarrow}\Phi(x) + h.c.) 
  \nonumber \\
  &+& \kappa|[\partial_\mu-i(a_\uparrow^\mu-a_\downarrow^\mu)]\Phi(x)|^2 \bigg{]}
  \label{Ssr2}
\end{eqnarray}
In the last line we include a kinetic term for $\Phi$, which minimally couples to 
$a_\uparrow^\mu-a_\downarrow^\mu$ to maintain gauge invariance.  

When $u$ exceeds a critical value, $\Phi$ condenses and the system enters
an interlayer coherent CFL phase.  To get a crude sense for when this transpires, one can integrate 
out the composite fermions to derive an effective theory for $\Phi$ coupled to $a_\uparrow^\mu-a_\downarrow^\mu$.  To leading order, the coefficient of the $|\Phi(x)|^2$ term shifts 
to $\frac{1}{u}-\frac{m^*}{2\pi}$.  Using our earlier estimate for $u$, this vanishes at a 
critical layer separation $(d/\ell_B)_c \approx e^2\ell_B m^*/2$.  Inserting Murthy and 
Shankar's estimate \cite{CFmass} $1/m^* \approx e^2 \ell_B/6$ yields $(d/\ell_B)_c \approx 3$.
Ultimately, however, $(d/\ell_B)_c$ should be determined numerically as we 
discuss below.

As an aside, we briefly comment on the case with $\nu_{\uparrow,\downarrow} = 1/4$, where the 
composite fermions have an effective filling $\nu^{CF}_{\uparrow,\downarrow} = 1/2$.  From 
Eqs.\ (\ref{Ssr}) and (\ref{Ssr2}) one similarly expects short-range interlayer repulsion to 
drive exciton condensation of composite fermions below a critical layer separation.    
Composite fermions then form the 111 state, so the electron wavefunction is 
$\psi = \prod_{i<j}(z_i-z_j)^2(w_i-w_j)^2\Psi_{111}$, \emph{i.e.}, the 331 state.  There is 
strong experimental \cite{Jim331,Shayegan331} and theoretical 
\cite{Theory331A,Theory331B,Theory331C} evidence that this phase indeed emerges at 
intermediate $d/\ell_B$.  Similarly, composite fermion exciton condensation 
at $\nu_{\uparrow,\downarrow} = 1/8$ (corresponding to $\nu^{CF}_{\uparrow,\downarrow} = 1/6$) generates the 553 state, which recent work \cite{Theory553} 
shows is a good candidate for the observed quantum Hall state at this 
filling \cite{OneQuarterPlateau}.  These observations substantiate the basic logic utilized above.

We now return to $\nu_{\uparrow,\downarrow} = 1/2$ and characterize the 
interlayer coherent CFL with $\langle \Phi \rangle \neq 0$.  The first remarkable property of 
this phase, which follows from Eq.\ (\ref{Ssr2}), is that composite fermions are liberated from 
their respective layers and coherently interlayer tunnel.  Consequently, `bonding' and 
`antibonding' composite Fermi surfaces form as shown in Fig.\ 
\ref{FermiSurfaces}(b).  The respective Fermi momenta $k_{F,B}$ and $k_{F,A}$ are 
determined by $d/\ell_B$ but must satisfy $k_{F,B}^2 + k_{F,A}^2 = 2/\ell_B^2$ to yield the 
correct filling factor.  In contrast, at $d/\ell_B = \infty$ composite fermions are confined 
to their layers and form independent, equal-size Fermi surfaces as 
in Fig.\ \ref{FermiSurfaces}(a).  

Crucially, although composite fermions tunnel the electrons do not (as in the 
331 state).  This can be understood by reformulating the problem using Wen's 
parton construction \cite{WenParton}, expressing the electron as 
$c_{{\bf r}\alpha} = b_{{\bf r}\alpha} f_{{\bf r}\alpha}$.  Here $b_{{\bf r}\alpha}$ is a boson 
that mimics the Chern-Simons flux attachment and $f_{{\bf r}\alpha}$ is the 
composite fermion; to remain in the physical Hilbert space one imposes the local constraint 
$c_{{\bf r}\alpha}^\dagger c_{{\bf r}\alpha} = b_{{\bf r}\alpha}^\dagger b_{{\bf r}\alpha} = f_{{\bf r}\alpha}^\dagger f_{{\bf r}\alpha}$.  
In the interlayer coherent CFL, the bosons form decoupled $\nu = 1/2$ quantum Hall states 
in each layer, while the composite fermions tunnel coherently.  Due to the constraint, electron 
tunneling requires \emph{both} $b_{{\bf r}\alpha}$ \emph{and} $f_{{\bf r}\alpha}$ to tunnel, but 
only the latter is able to do so as Fig.\ \ref{BilayerFig} illustrates.  

\begin{figure}
\centering{
  \includegraphics[width=3in]{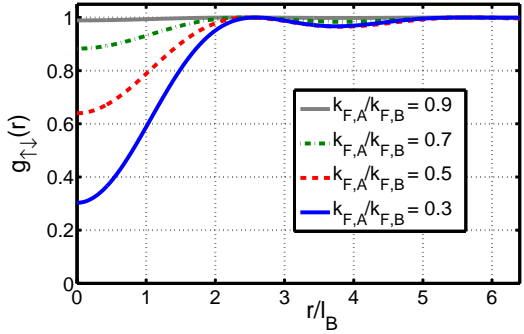}
  \caption{(Color online) Interlayer density-density correlation function in the interlayer coherent CFL 
  computed in mean-field theory for various $k_{F,A}/k_{F,B}$.  Reducing $k_{F,A}/k_{F,B}$ smoothly carves out an interlayer correlation hole as seen above.  }
  \label{PairCorrelationFig}}
\end{figure}

Partons also allow correlations to be simply computed in a 
mean-field approximation that neglects the Hilbert-space constraint.  
Consider the interlayer density-density correlation function 
$g_{\uparrow\downarrow}({\bf r}-{\bf r}') = \rho^{-2}\langle c_{{\bf r}\uparrow}^\dagger c_{{\bf r'}\downarrow}^\dagger c_{{\bf r'}\downarrow}c_{{\bf r}\uparrow}\rangle$, where $\rho = 1/(4\pi\ell_B^2)$. 
Upon introducing partons and ignoring the constraint, 
$g_{\uparrow\downarrow}$ factorizes: $g_{\uparrow\downarrow}({\bf r}-{\bf r}') = \rho^{-2}\langle b_{{\bf r}\uparrow}^\dagger b_{{\bf r'}\downarrow}^\dagger b_{{\bf r'}\downarrow}b_{{\bf r}\uparrow}\rangle\langle f_{{\bf r}\uparrow}^\dagger f_{{\bf r'}\downarrow}^\dagger f_{{\bf r'}\downarrow}f_{{\bf r}\uparrow}\rangle$.  
The boson part yields a constant since the exchange term vanishes.  Setting this 
contribution to unity (\emph{i.e.}, using `renormalized mean field theory') 
yields $g_{\uparrow\downarrow}({\bf r}-{\bf r}') \rightarrow \rho^{-2}\langle f_{{\bf r}\uparrow}^\dagger f_{{\bf r'}\downarrow}^\dagger f_{{\bf r'}\downarrow}f_{{\bf r}\uparrow}\rangle$, 
which evaluates to
\begin{eqnarray}
  g_{\uparrow\downarrow}({\bf r}) &=& 1-\frac{\ell_B^4}{r^2}[k_{F,B}J_1(k_{F,B} r)-k_{F,A}J_1(k_{F,A} r)]^2.
\end{eqnarray}
Figure \ref{PairCorrelationFig} displays $g_{\uparrow\downarrow}$ for several values 
of $k_{F,A}/k_{F,B}$, and demonstrates that reducing $k_{F,A}/k_{F,B}$ lowers the interlayer Coulomb energy by smoothly binding an interlayer correlation hole.   
Interestingly, this crude treatment already captures 
the oscillations in $g_{\uparrow\downarrow}$ found numerically in exact diagonalization (Fig.\ 10(b) 
in \cite{Moller}) and DMRG \cite{BilayerDMRG}. 
 
In contrast to the electron exciton condensation at
small $d/\ell_B$ which spontaneously breaks a physical $U(1)$ symmetry
(corresponding to electron number conservation in each layer), 
the composite fermion exciton condensate 
only breaks gauge symmetry.  This follows from Eq.\ (\ref{Ssr2}), where condensing $\Phi$ yields a mass term 
for $a_\uparrow^\mu-a_\downarrow^\mu$, thereby pinning the Chern-Simons fields for the two 
layers together.  Interesting physical consequences follow, which we now discuss.  

Electromagnetic response properties are most clearly organized in a basis of 
symmetric/antisymmetric currents $j^\mu_{s/a} = \frac{1}{\sqrt{2}}(j^\mu_{\uparrow} \pm j^\mu_\downarrow)$.  The response of $j^0_{s/a}$ yields the compressibility with respect to 
symmetric density changes ($\nu_{\uparrow,\downarrow} = 1/2 + \delta\nu$), and layer 
imbalance ($\nu_{\uparrow} = 1/2 + \delta\nu$, $\nu_\downarrow = 1/2-\delta\nu$).  Spatial 
components ${\vec j}_{s/a}$, corresponding to symmetric and counterflow currents, respond to 
electric fields through ${\vec E}_{s/a} = \overleftrightarrow{\rho}_{s/a}{\vec j}_{s/a}$, with 
${\vec E}_{s/a} = \frac{1}{\sqrt{2}}({\vec E}_\uparrow \pm {\vec E}_{\downarrow})$.  To evaluate the 
$q\rightarrow 0$, static compressibilities and resistivities, it suffices to simply set 
$a_\uparrow = a_\downarrow \equiv a$ since $a_\uparrow-a_\downarrow$ is massive.  The composite 
fermion action can then be written
\begin{eqnarray}
  S_{CF} &\rightarrow & \int_{x}\sum_{\alpha = \uparrow,\downarrow} f_{x\alpha}^\dagger \bigg{[}(\partial_\tau-ia^0)  - \frac{(\nabla -i {\bf a})^2}{2m^*}\bigg{]} f_{x\alpha} 
  \nonumber \\
  &-&t[f_{x\uparrow}^\dagger f_{x\downarrow} + h.c.] + S_{Coul}^{\uparrow\uparrow} + S_{Coul}^{\downarrow\downarrow} + S_{lr}^{\uparrow\downarrow}
  \nonumber \\
  &+& \int_{x}\bigg{[}\frac{i}{8\pi}(\sqrt{2}a^\mu+e\mathcal{A}_s^\mu) \epsilon_{\mu\nu\lambda}\partial_\nu (\sqrt{2}a^\lambda + e\mathcal{A}^\lambda_s) 
  \nonumber \\
  &+& \frac{ie^2}{8\pi}\mathcal{A}_a^\mu \epsilon_{\mu\nu\lambda}\partial_\nu \mathcal{A}^\lambda_a \bigg{]},
  \label{SCFpinned}
\end{eqnarray}
with $t = \langle \Phi \rangle$ taken real and 
$\mathcal{A}^\mu_{s/a} = \frac{1}{\sqrt{2}}(\mathcal{A}^\mu_{\uparrow} \pm \mathcal{A}^\mu_\downarrow)$.  

The Chern-Simons term for $\mathcal{A}_{a}$, which decouples from 
everything else, is the effective action for a $\nu = 1/2$ fractional quantum Hall state. 
This has remarkable implications: despite having two Fermi surfaces, \emph{the interlayer 
coherent CFL behaves like an incompressible, gapped fractional quantum Hall state in the 
counterflow channel}, with resistivity $\rho_a^{xx} = 0$ and $\rho_a^{xy} = 2h/e^2$.  
As noted by Stern and Halperin \cite{GaugeInvarianceCompressibility}, compressibility of a CFL 
is intimately tied to gauge invariance.  Incompressibility thus formally stems from the 
partial breaking of gauge symmetry due to $\Phi$ condensing.  More physically, transferring 
electrons from one layer to the other requires creating a net difference in flux between the 
layers; with $a_\uparrow-a_\downarrow$ massive doing so requires overcoming an energy gap.  

To study such charge excitations, consider the more general composite fermion action, 
Eqs.\ (\ref{SCF}) and (\ref{Ssr2}).  Elementary charge excitations in this channel are created 
by inserting a vortex in $\Phi$, which by a singular gauge transformation is equivalent to 
adding localized $\pi$ flux in $a_\uparrow$ and $-\pi$ flux in $a_\downarrow$.  A 
physical electron consists of a composite fermion bound to $4\pi$ flux, so the quasiparticles are formed 
by dipoles carrying charge $e/4$ in one layer and $-e/4$ in the other.  

Since gauge symmetry is preserved in the symmetric channel, the interlayer coherent CFL and 
decoupled CFL's behave essentially identically here.  Both are compressible, and have 
resistivity elements 
$\rho_s^{xy} = 2h/e^2$ and $\rho_s^{xx} = \rho^{xx}_{CF}$, where $\rho^{xx}_{CF}$ is the composite 
fermion resistivity with disorder.  This result can be obtained via the methods of \cite{HLR}, or in the parton approach using the Ioffe-Larkin rule \cite{IoffeLarkin} which states that ${\rho}_s^{ij}$ is the sum of resistivities for the bosons, ${\rho}_b^{ij} = (2h/e^2)\epsilon_{ij}$, and composite fermions, ${\rho}^{ij}_{CF} = \rho_{CF}^{xx}\delta_{ij}$.  

Given the compressibility in this channel, it is interesting to ask how the system evolves when 
the field shifts to $B = B_{1/2} \pm |\delta B|$ so that 
$\nu_{\uparrow,\downarrow} = 1/2 \mp |\delta\nu|$.  
The attached flux now only partially cancels the field, and the bonding/antibonding Fermi 
surfaces develop into Landau levels such that the effective composite fermion filling 
is $\nu^{CF}_{\uparrow,\downarrow} = 2\pi\rho/(e|\delta B|) = \nu_{\uparrow,\downarrow}/(2|\delta\nu|)$.
Incompressible phases arise whenever an integer number $p_{B/A}$ of bonding/antibonding Landau 
levels are filled---\emph{i.e.}, when $\nu^{CF}_{\uparrow,\downarrow} = (p_B+p_A)/2$.  
The corresponding electron filling factors and LLL-projected wavefunctions are
\begin{eqnarray}
  \nu_{\uparrow,\downarrow} &=& \frac{p_A + p_B}{2(p_A+p_B \pm 1)}
  \label{nu}
  \\
  \psi_{(p_B,p_A)} &=& P_{LLL}\prod_{i<j}(z_i-z_j)^2(w_i-w_j)^2\Psi_{(p_B,p_A)}
  \label{psipApB}
\end{eqnarray}
where $\Psi_{(p_B,p_A)}$ is the wavefunction for $p_{B/A}$ filled bonding/antibonding composite 
fermion Landau levels.  This series of interlayer-correlated quantum Hall states constitutes a 
bilayer analogue of the Jain sequence \cite{JainCF}, and notably includes 
\emph{even denominator} fractions.  Just as the Jain sequence can be viewed as quantum oscillations of 
a single-layer CFL \cite{HLR}, this bilayer series emerges naturally as 
quantum oscillations of the interlayer coherent CFL; consequently, appearance of these fractions 
can serve as indirect evidence for its existence.  

The properties of the interlayer coherent CFL discussed 
above readily distinguish it from other proposals for intermediate $d/\ell_B$ 
at $\nu_{\uparrow,\downarrow} = 1/2$.  For example, the interlayer BCS paired state is incompressible 
in both the symmetric and antisymmetric channels \cite{DipoleBCSpicture}.  Compared to the latter, 
the interlayer coherent CFL bears numerical advantages, since the trial wavefunction 
in Eq.\ (\ref{coherentCFLpsi}) has one variational 
parameter---$k_{F,A}/k_{F,B}$---whereas BCS states require variational determination of the 
pairing \emph{function} \cite{MollerPRL,Moller}.  Thus, variational Monte Carlo can be employed for 
relatively large particle numbers to estimate more seriously $(d/\ell_B)_c$ below which 
composite fermions exciton condense, and to study the optimal Fermi surface sizes at smaller 
$d/\ell_B$.  Comparison with exact diagonalization for smaller systems may also be 
fruitful.  

Experimentally, in double quantum wells studied to date it seems unlikely that an 
interlayer coherent CFL is operative---a biproduct of the counterflow incompressibility is a 
large longitudinal drag $\sim \rho_{CF}^{xx}$, which is not observed \cite{JimDrag}.  This could be 
due to disorder-induced local filling factor variations, which this
phase strongly disfavors, or perhaps partial spin polarization \cite{JimSpinPolarization}. 
Wide quantum wells appear more promising: significantly larger mobilities are 
achievable, and quantum Hall states have been observed at several of the fractions predicted by 
Eq.\ (\ref{nu}) \cite{Shayegan, TsuiPhysicaE,OneQuarterPlateau}.  Further studies of 
quantum oscillations of the compressible phase near $\nu_{\uparrow,\downarrow} = 1/2$ would be 
exciting to more directly compare with our predictions, as would experiments to measure the 
unequal Fermi surfaces that are a hallmark of this state \cite{CFbook}.  Theoretically, composite fermions emerging as delocalized degrees of freedom in multilayers is a novel possibility that may open new avenues in quantum Hall physics.  A three-dimensional version of the bilayer coherent CFL  
---possibly relevant for layered semimetals like graphite---will be explored in future 
work.

\acknowledgments{It is a pleasure to acknowledge illuminating discussions with G.\ Chen, 
J.\ Eisenstein, Y.-B.\ Kim, D. Luhman, M.\ Milovanovic, G.\ Moller, and Y. Zou, as well as the hospitality of the Max Planck Institute for the Physics of Complex Systems where part of this work was carried out.  
We also acknowledge support from the 
Lee A.\ DuBridge Foundation (JA), the A.\ P.\ Sloan Foundation (OIM), the Packard Foundation (GR), and the National Science Foundation through grants 
DMR-0529399 (MPAF), PHY-0456720 and PHY-0803371 (GR).


\begin{thebibliography}{26}
\expandafter\ifx\csname natexlab\endcsname\relax\def\natexlab#1{#1}\fi
\expandafter\ifx\csname bibnamefont\endcsname\relax
  \def\bibnamefont#1{#1}\fi
\expandafter\ifx\csname bibfnamefont\endcsname\relax
  \def\bibfnamefont#1{#1}\fi
\expandafter\ifx\csname citenamefont\endcsname\relax
  \def\citenamefont#1{#1}\fi
\expandafter\ifx\csname url\endcsname\relax
  \def\url#1{\texttt{#1}}\fi
\expandafter\ifx\csname urlprefix\endcsname\relax\def\urlprefix{URL }\fi
\providecommand{\bibinfo}[2]{#2}
\providecommand{\eprint}[2][]{\url{#2}}

\bibitem[{\citenamefont{Jain}(1989)}]{JainCF}
\bibinfo{author}{\bibfnamefont{J.~K.} \bibnamefont{Jain}},
  \bibinfo{journal}{Phys.\ Rev.\ Lett.} \textbf{\bibinfo{volume}{63}},
  \bibinfo{pages}{199} (\bibinfo{year}{1989}).

\bibitem[{\citenamefont{Halperin et~al.}(1993)\citenamefont{Halperin, Lee, and
  Read}}]{HLR}
\bibinfo{author}{\bibfnamefont{B.~I.} \bibnamefont{Halperin}},
  \bibinfo{author}{\bibfnamefont{P.~A.} \bibnamefont{Lee}}, \bibnamefont{and}
  \bibinfo{author}{\bibfnamefont{N.}~\bibnamefont{Read}},
  \bibinfo{journal}{Phys.\ Rev.\ B} \textbf{\bibinfo{volume}{47}},
  \bibinfo{pages}{7312} (\bibinfo{year}{1993}).

\bibitem[{\citenamefont{Heinonen}(1998)}]{CFbook}
\bibinfo{editor}{\bibfnamefont{O.}~\bibnamefont{Heinonen}}, ed.,
  \emph{\bibinfo{title}{Composite Fermions: A Unified View of the Quantum Hall
  Regime}} (\bibinfo{publisher}{World Scientific, Singapore},
  \bibinfo{year}{1998}).

\bibitem[{\citenamefont{Kellogg et~al.}(2003)\citenamefont{Kellogg, Eisenstein,
  Pfeiffer, and West}}]{JimDrag}
\bibinfo{author}{\bibfnamefont{M.}~\bibnamefont{Kellogg}},
  \bibinfo{author}{\bibfnamefont{J.~P.} \bibnamefont{Eisenstein}},
  \bibinfo{author}{\bibfnamefont{L.~N.} \bibnamefont{Pfeiffer}},
  \bibnamefont{and} \bibinfo{author}{\bibfnamefont{K.~W.} \bibnamefont{West}},
  \bibinfo{journal}{Phys.\ Rev.\ Lett.} \textbf{\bibinfo{volume}{90}},
  \bibinfo{pages}{246801} (\bibinfo{year}{2003}).

\bibitem[{\citenamefont{Behnia et~al.}(2007)\citenamefont{Behnia, Balicas, and
  Kopelevich}}]{Behnia}
\bibinfo{author}{\bibfnamefont{K.}~\bibnamefont{Behnia}},
  \bibinfo{author}{\bibfnamefont{L.}~\bibnamefont{Balicas}}, \bibnamefont{and}
  \bibinfo{author}{\bibfnamefont{Y.}~\bibnamefont{Kopelevich}},
  \bibinfo{journal}{Science} \textbf{\bibinfo{volume}{317}},
  \bibinfo{pages}{1729} (\bibinfo{year}{2007}).

\bibitem[{\citenamefont{Das~Sarma and Pinczuk}(1997)}]{QHbook}
\bibinfo{editor}{\bibfnamefont{S.}~\bibnamefont{Das~Sarma}} \bibnamefont{and}
  \bibinfo{editor}{\bibfnamefont{A.}~\bibnamefont{Pinczuk}}, eds.,
  \emph{\bibinfo{title}{Perspectives in Quantum Hall Effects}}
  (\bibinfo{publisher}{Wiley, New York}, \bibinfo{year}{1997}).

\bibitem[{\citenamefont{M\"{o}ller et~al.}(2009)\citenamefont{M\"{o}ller,
  Simon, and Rezayi}}]{Moller}
\bibinfo{author}{\bibfnamefont{G.}~\bibnamefont{M\"{o}ller}},
  \bibinfo{author}{\bibfnamefont{S.~H.} \bibnamefont{Simon}}, \bibnamefont{and}
  \bibinfo{author}{\bibfnamefont{E.~H.} \bibnamefont{Rezayi}},
  \bibinfo{journal}{Phys.\ Rev.\ B} \textbf{\bibinfo{volume}{79}},
  \bibinfo{pages}{125106} (\bibinfo{year}{2009}).

\bibitem[{\citenamefont{Halperin}(1984)}]{mmn}
\bibinfo{author}{\bibfnamefont{B.~I.} \bibnamefont{Halperin}},
  \bibinfo{journal}{Helv.\ Phys.\ Acta.} \textbf{\bibinfo{volume}{56}},
  \bibinfo{pages}{75} (\bibinfo{year}{1984}).

\bibitem[{\citenamefont{Bonesteel et~al.}(1996)\citenamefont{Bonesteel,
  McDonald, and Nayak}}]{BonesteelNayak}
\bibinfo{author}{\bibfnamefont{N.~E.} \bibnamefont{Bonesteel}},
  \bibinfo{author}{\bibfnamefont{I.~A.} \bibnamefont{McDonald}},
  \bibnamefont{and} \bibinfo{author}{\bibfnamefont{C.}~\bibnamefont{Nayak}},
  \bibinfo{journal}{Phys.\ Rev.\ Lett.} \textbf{\bibinfo{volume}{77}},
  \bibinfo{pages}{3009} (\bibinfo{year}{1996}).

\bibitem[{\citenamefont{Kim et~al.}(2001)\citenamefont{Kim, Nayak, Demler,
  Read, and Das~Sarma}}]{DipoleBCSpicture}
\bibinfo{author}{\bibfnamefont{Y.~B.} \bibnamefont{Kim}},
  \bibinfo{author}{\bibfnamefont{C.}~\bibnamefont{Nayak}},
  \bibinfo{author}{\bibfnamefont{E.}~\bibnamefont{Demler}},
  \bibinfo{author}{\bibfnamefont{N.}~\bibnamefont{Read}}, \bibnamefont{and}
  \bibinfo{author}{\bibfnamefont{S.}~\bibnamefont{Das~Sarma}},
  \bibinfo{journal}{Phys.\ Rev.\ B} \textbf{\bibinfo{volume}{63}},
  \bibinfo{pages}{205315} (\bibinfo{year}{2001}).

\bibitem[{\citenamefont{Shankar and Murthy}(1997)}]{CFmass}
\bibinfo{author}{\bibfnamefont{R.}~\bibnamefont{Shankar}} \bibnamefont{and}
  \bibinfo{author}{\bibfnamefont{G.}~\bibnamefont{Murthy}},
  \bibinfo{journal}{Phys.\ Rev.\ Lett.} \textbf{\bibinfo{volume}{79}},
  \bibinfo{pages}{4437} (\bibinfo{year}{1997}).

\bibitem[{\citenamefont{Eisenstein et~al.}(1992)\citenamefont{Eisenstein,
  Boebinger, Pfeiffer, West, and He}}]{Jim331}
\bibinfo{author}{\bibfnamefont{J.~P.} \bibnamefont{Eisenstein}},
  \bibinfo{author}{\bibfnamefont{G.~S.} \bibnamefont{Boebinger}},
  \bibinfo{author}{\bibfnamefont{L.~N.} \bibnamefont{Pfeiffer}},
  \bibinfo{author}{\bibfnamefont{K.~W.} \bibnamefont{West}}, \bibnamefont{and}
  \bibinfo{author}{\bibfnamefont{S.}~\bibnamefont{He}},
  \bibinfo{journal}{Phys.\ Rev.\ Lett.} \textbf{\bibinfo{volume}{68}},
  \bibinfo{pages}{1383} (\bibinfo{year}{1992}).

\bibitem[{\citenamefont{Suen et~al.}(1994)\citenamefont{Suen, Manoharan, Ying,
  Santos, and Shayegan}}]{Shayegan331}
\bibinfo{author}{\bibfnamefont{Y.~W.} \bibnamefont{Suen}},
  \bibinfo{author}{\bibfnamefont{H.~C.} \bibnamefont{Manoharan}},
  \bibinfo{author}{\bibfnamefont{X.}~\bibnamefont{Ying}},
  \bibinfo{author}{\bibfnamefont{M.~B.} \bibnamefont{Santos}},
  \bibnamefont{and} \bibinfo{author}{\bibfnamefont{M.}~\bibnamefont{Shayegan}},
  \bibinfo{journal}{Phys.\ Rev.\ Lett.} \textbf{\bibinfo{volume}{72}},
  \bibinfo{pages}{3405} (\bibinfo{year}{1994}).

\bibitem[{\citenamefont{Yoshioka et~al.}(1989)\citenamefont{Yoshioka,
  MacDonald, and Girvin}}]{Theory331A}
\bibinfo{author}{\bibfnamefont{D.}~\bibnamefont{Yoshioka}},
  \bibinfo{author}{\bibfnamefont{A.~H.} \bibnamefont{MacDonald}},
  \bibnamefont{and} \bibinfo{author}{\bibfnamefont{S.~M.}
  \bibnamefont{Girvin}}, \bibinfo{journal}{Phys.\ Rev.\ B}
  \textbf{\bibinfo{volume}{39}}, \bibinfo{pages}{1932} (\bibinfo{year}{1989}).

\bibitem[{\citenamefont{He et~al.}(1991)\citenamefont{He, Xie, Das~Sarma, and
  Zhang}}]{Theory331B}
\bibinfo{author}{\bibfnamefont{S.}~\bibnamefont{He}},
  \bibinfo{author}{\bibfnamefont{X.~C.} \bibnamefont{Xie}},
  \bibinfo{author}{\bibfnamefont{S.}~\bibnamefont{Das~Sarma}},
  \bibnamefont{and} \bibinfo{author}{\bibfnamefont{F.~C.} \bibnamefont{Zhang}},
  \bibinfo{journal}{Phys.\ Rev.\ B} \textbf{\bibinfo{volume}{43}},
  \bibinfo{pages}{9339} (\bibinfo{year}{1991}).

\bibitem[{\citenamefont{Nomura and Yoshioka}(2004)}]{Theory331C}
\bibinfo{author}{\bibfnamefont{K.}~\bibnamefont{Nomura}} \bibnamefont{and}
  \bibinfo{author}{\bibfnamefont{D.}~\bibnamefont{Yoshioka}},
  \bibinfo{journal}{J.\ Phys.\ Soc.\ Jpn.} \textbf{\bibinfo{volume}{73}},
  \bibinfo{pages}{2612} (\bibinfo{year}{2004}).

\bibitem[{\citenamefont{Papi\'{c} et~al.}(2009)\citenamefont{Papi\'{c},
  M\"{o}ller, Milovanovi\'{c}, Regnault, and Goerbig}}]{Theory553}
\bibinfo{author}{\bibfnamefont{Z.}~\bibnamefont{Papi\'{c}}},
  \bibinfo{author}{\bibfnamefont{G.}~\bibnamefont{M\"{o}ller}},
  \bibinfo{author}{\bibfnamefont{M.~V.} \bibnamefont{Milovanovi\'{c}}},
  \bibinfo{author}{\bibfnamefont{N.}~\bibnamefont{Regnault}}, \bibnamefont{and}
  \bibinfo{author}{\bibfnamefont{M.~O.} \bibnamefont{Goerbig}},
  \bibinfo{journal}{Phys.\ Rev.\ B} \textbf{\bibinfo{volume}{79}},
  \bibinfo{pages}{245325} (\bibinfo{year}{2009}).

\bibitem[{\citenamefont{Luhman et~al.}(2008{\natexlab{a}})\citenamefont{Luhman,
  Pan, Tsui, Pfeiffer, Baldwin, and West}}]{OneQuarterPlateau}
\bibinfo{author}{\bibfnamefont{D.~R.} \bibnamefont{Luhman}},
  \bibinfo{author}{\bibfnamefont{W.}~\bibnamefont{Pan}},
  \bibinfo{author}{\bibfnamefont{D.~C.} \bibnamefont{Tsui}},
  \bibinfo{author}{\bibfnamefont{L.~N.} \bibnamefont{Pfeiffer}},
  \bibinfo{author}{\bibfnamefont{K.~W.} \bibnamefont{Baldwin}},
  \bibnamefont{and} \bibinfo{author}{\bibfnamefont{K.~W.} \bibnamefont{West}},
  \bibinfo{journal}{Phys.\ Rev.\ Lett.} \textbf{\bibinfo{volume}{101}},
  \bibinfo{pages}{266804} (\bibinfo{year}{2008}{\natexlab{a}}).

\bibitem[{\citenamefont{Wen}(1999)}]{WenParton}
\bibinfo{author}{\bibfnamefont{X.-G.} \bibnamefont{Wen}},
  \bibinfo{journal}{Phys.\ Rev.\ B} \textbf{\bibinfo{volume}{60}},
  \bibinfo{pages}{8827} (\bibinfo{year}{1999}).

\bibitem[{\citenamefont{Shibata and Yoshioka}(2006)}]{BilayerDMRG}
\bibinfo{author}{\bibfnamefont{N.}~\bibnamefont{Shibata}} \bibnamefont{and}
  \bibinfo{author}{\bibfnamefont{D.}~\bibnamefont{Yoshioka}},
  \bibinfo{journal}{J.\ Phys.\ Soc.\ Jpn.} \textbf{\bibinfo{volume}{75}},
  \bibinfo{pages}{043712} (\bibinfo{year}{2006}).

\bibitem[{\citenamefont{Halperin and
  Stern}(1998)}]{GaugeInvarianceCompressibility}
\bibinfo{author}{\bibfnamefont{B.~I.} \bibnamefont{Halperin}} \bibnamefont{and}
  \bibinfo{author}{\bibfnamefont{A.}~\bibnamefont{Stern}},
  \bibinfo{journal}{Phys.\ Rev.\ Lett.} \textbf{\bibinfo{volume}{80}},
  \bibinfo{pages}{5457} (\bibinfo{year}{1998}).

\bibitem[{\citenamefont{Ioffe and Larkin}(1989)}]{IoffeLarkin}
\bibinfo{author}{\bibfnamefont{L.~B.} \bibnamefont{Ioffe}} \bibnamefont{and}
  \bibinfo{author}{\bibfnamefont{A.~I.} \bibnamefont{Larkin}},
  \bibinfo{journal}{Phys.\ Rev.\ B} \textbf{\bibinfo{volume}{39}},
  \bibinfo{pages}{8988} (\bibinfo{year}{1989}).

\bibitem[{\citenamefont{M\"{o}ller et~al.}(2008)\citenamefont{M\"{o}ller,
  Simon, and Rezayi}}]{MollerPRL}
\bibinfo{author}{\bibfnamefont{G.}~\bibnamefont{M\"{o}ller}},
  \bibinfo{author}{\bibfnamefont{S.~H.} \bibnamefont{Simon}}, \bibnamefont{and}
  \bibinfo{author}{\bibfnamefont{E.~H.} \bibnamefont{Rezayi}},
  \bibinfo{journal}{Phys.\ Rev.\ Lett.} \textbf{\bibinfo{volume}{101}},
  \bibinfo{pages}{176803} (\bibinfo{year}{2008}).

\bibitem[{\citenamefont{Spielman et~al.}(2005)\citenamefont{Spielman, Tracy,
  Eisenstein, Pfeiffer, and West}}]{JimSpinPolarization}
\bibinfo{author}{\bibfnamefont{I.~B.} \bibnamefont{Spielman}},
  \bibinfo{author}{\bibfnamefont{L.~A.} \bibnamefont{Tracy}},
  \bibinfo{author}{\bibfnamefont{J.~P.} \bibnamefont{Eisenstein}},
  \bibinfo{author}{\bibfnamefont{L.~N.} \bibnamefont{Pfeiffer}},
  \bibnamefont{and} \bibinfo{author}{\bibfnamefont{K.~W.} \bibnamefont{West}},
  \bibinfo{journal}{Phys.\ Rev.\ Lett.} \textbf{\bibinfo{volume}{94}},
  \bibinfo{pages}{076803} (\bibinfo{year}{2005}).

\bibitem[{\citenamefont{Shayegan et~al.}(1996)\citenamefont{Shayegan,
  Manoharan, Suen, Lay, and Santos}}]{Shayegan}
\bibinfo{author}{\bibfnamefont{M.}~\bibnamefont{Shayegan}},
  \bibinfo{author}{\bibfnamefont{H.~C.} \bibnamefont{Manoharan}},
  \bibinfo{author}{\bibfnamefont{Y.~W.} \bibnamefont{Suen}},
  \bibinfo{author}{\bibfnamefont{T.~S.} \bibnamefont{Lay}}, \bibnamefont{and}
  \bibinfo{author}{\bibfnamefont{M.~B.} \bibnamefont{Santos}},
  \bibinfo{journal}{Semicond.\ Sci.\ Technol.} \textbf{\bibinfo{volume}{11}},
  \bibinfo{pages}{1539} (\bibinfo{year}{1996}).

\bibitem[{\citenamefont{Luhman et~al.}(2008{\natexlab{b}})\citenamefont{Luhman,
  Pan, Tsui, Pfeiffer, Baldwin, and West}}]{TsuiPhysicaE}
\bibinfo{author}{\bibfnamefont{D.}~\bibnamefont{Luhman}},
  \bibinfo{author}{\bibfnamefont{W.}~\bibnamefont{Pan}},
  \bibinfo{author}{\bibfnamefont{D.}~\bibnamefont{Tsui}},
  \bibinfo{author}{\bibfnamefont{L.}~\bibnamefont{Pfeiffer}},
  \bibinfo{author}{\bibfnamefont{K.}~\bibnamefont{Baldwin}}, \bibnamefont{and}
  \bibinfo{author}{\bibfnamefont{K.}~\bibnamefont{West}},
  \bibinfo{journal}{Physica E} \textbf{\bibinfo{volume}{40}},
  \bibinfo{pages}{1059} (\bibinfo{year}{2008}{\natexlab{b}}).

\end{thebibliography}

\end{document}